\PassOptionsToPackage{hyphens}{url}
\documentclass[runningheads]{llncs}
\usepackage[T1]{fontenc}
\usepackage{graphicx}
\usepackage{listings}
\usepackage{booktabs}
\usepackage{bbding}
\usepackage{url}
\usepackage{hyperref}
\hypersetup{breaklinks=true}

\usepackage{orcidlink}

\begin{document}

\title{LLM-Based Information Extraction to Support Scientific Literature
       Research and Publication Workflows \thanks{Published in TPDL 2025, New Trends in Theory and Practice of Digital Libraries, Communications in Computer and Information Science, vol 2694. DOI \href{https:/doi.org/10.1007/978-3-032-06136-2_9}{10.1007/978-3-032-06136-2\_9}. This PDF is the author-prepared camera-ready version corresponding to the accepted manuscript and supersedes the submitted version that was inadvertently published as the version of record.}}

\titlerunning{LLM Info Extraction for Research Workflows}
\author{Samy Ateia\inst{1}\orcidlink{0009-0000-2622-9194}\Envelope \and
Udo Kruschwitz\inst{1}\orcidlink{0000-0002-5503-0341} \and
Melanie Scholz\inst{2} \and
Agnes Koschmider\inst{2}\orcidlink{0000-0001-8206-7636} \and
Moayad Almohaishi\inst{2}\orcidlink{0009-0002-1758-3153}
}

\authorrunning{S. Ateia et al.}

\institute{University of Regensburg, Universitätsstraße 31, 93053 Regensburg, Germany\\
\email{\{udo.kruschwitz,samy.ateia\}@ur.de}\\ \and
University of Bayreuth, Universitätsstraße 30, 95447 Bayreuth, Germany\\
\email{\{melanie.scholz,agnes.koschmider,moayad.almohaishi\}@uni-bayreuth.de}}  

\maketitle

\begin{abstract}  
The increasing volume of scholarly publications requires advanced tools for efficient knowledge discovery and management. This paper introduces ongoing work on a system using Large Language Models (LLMs) for the semantic extraction of key concepts from scientific documents. Our research, conducted within the German National Research Data Infrastructure for and with Computer Science (NFDIxCS) project, seeks to support FAIR (Findable, Accessible, Interoperable, and Reusable) principles in scientific publishing. We outline our explorative work, which uses in-context learning with various LLMs to extract concepts from papers, initially focusing on the Business Process Management (BPM) domain. A key advantage of this approach is its potential for rapid domain adaptation, often requiring few or even zero examples to define extraction targets for new scientific fields. We conducted technical evaluations to compare the performance of commercial and open-source LLMs and created an online demo application to collect feedback from an initial user-study. Additionally, we gathered insights from the computer science research community through user stories collected during a dedicated workshop, actively guiding the ongoing development of our future services. These services aim to support structured literature reviews, concept-based information retrieval, and integration of extracted knowledge into existing knowledge graphs.

\end{abstract}

\section{Introduction}
The scientific publication landscape is booming with global annual publication growing by 59\%  according to the NSF\footnote{\url{https://web.archive.org/web/20250507134337/https://www.ncses.nsf.gov/pubs/nsb202333/executive-summary}} and more than one million articles being published per year in biomedicine and life sciences alone \cite{GONZALEZMARQUEZ2024100968}. This rise in publications makes it harder for scientists to stay on top of their field, while also limiting the discoverability of their publications as they have to compete with others for visibility. Digital transformation and new tools such as LLMs can accelerate that problem, but also have the potential to assist scientists and publishing platforms in managing these challenges.

In computer science, publications are often accompanied by software artifacts and datasets for reproducibility, but their management frequently lacks standardization and fails to meet FAIR principles. The National Research Data Infrastructure for and with Computer Science (NFDIxCS) project\footnote{\url{https://nfdixcs.org/}} addresses this by creating an infrastructure to implement FAIR principles \cite{wilkinson2016fair} for CS research outputs in Germany \cite{goedicke2022research}. But to make these artifacts findable it is necessary to link them to relevant semantic information from the publication text itself, for example, research questions or methods so that they are discoverably as related work by other scientists.

Our project, situated within the NFDIxCS initiative, aims to develop tools to exactly address this problem. We leverage Large Language Models (LLMs) for the semantic analysis of scientific text, with the goal of enhancing the FAIR principles for scholarly literature. Specifically, we aim to:
\begin{itemize}
    \item Develop robust methods for automatically extracting key semantic concepts (e.g., research questions, methodologies, findings) from scientific papers.
    \item Explore mechanisms for structuring these extracted concepts to improve the organization and distribution of digital content, potentially linking them to knowledge graphs.
    \item Design and prototype services, informed by community needs, that use this structured information to support researchers in their workflows.
\end{itemize}
This short paper presents our preliminary findings and outlines how user-driven requirements are shaping the trajectory of our research towards practical applications. We publish a demo system alongside its source code under a permissive license\footnote{CC BY 4.0} alongside the results of our user workshops, and plan to maintain this practice for future services.

\section{Related Work}
In this work, we explore the use of LLMs for knowledge extraction to improve scientific workflows within digital libraries and beyond. We review related efforts in three key areas: platforms for scientific literature analysis, knowledge graphs for structuring scientific information, and the use of LLMs for information extraction.

% I am thinking here that one big disadvantage of Elicit and Scite.ai is the paid service, Elicit limited to 20 papers ( nearly two analsis reports) and Scite.ai gives a 7 day tryout on an on-line trial. which in my opinion, does not help in Promoting FAIR research.
\subsection{Platforms for Scientific Literature Analysis}
Several platforms exist that use natural language processing (NLP) based on language models to highlight relevant information from scientific text, therefore assisting in navigating the considerable volume of publications. \textbf{Semantic Scholar} \cite{ammar2018construction}, for example, employs AI to provide summaries (TLDRs) and identify influential citations, while \textbf{Elicit} uses a systematic review inspired workflow, leveraging LLMs to synthesize findings from multiple papers in response to a user's query \cite{whitfield2023elicit}. \textbf{Scite.ai} focuses specifically on citation context, classifying whether a citation supports, disputes, or merely mentions a claim \cite{lund2023examining}. While these platforms offer similar flexible LLM-based question answering tools, they do not currently offer the use of predefined domain-specific questions and mostly require a paid subscription to be fully utilized. With our approach, we want to potentially offer higher accuracy and user guidance through curated extraction targets and examples serving specialized communities.

\subsection{Knowledge Graphs for Structuring Scientific Knowledge}
Structuring scientific knowledge in a machine-readable format has long been a goal of the scientific community. The Open Research Knowledge Graph (ORKG) \cite{jaradeh2019open} is a prominent initiative aiming to represent the content of research papers as structured data. By describing papers through their contributions, methods, and findings, the ORKG facilitates systematic comparisons and reviews. Other notable examples are the discontinued Microsoft Academic Graph (MAG) \cite{wang2020microsoft} that was succeeded by OpenAlex \cite{priem2022openalex} or SciKGraph \cite{TOSI2021101109}.
However, curating such knowledge graphs often requires significant manual effort from researchers. Most recently, ORKG Ask was introduced, which offers the possibility to create ad-hoc comparison tables using information extraction with LLMs \cite{oelen2024orkg}.
In our work, we want to build on that approach and take it a step further. Instead of just having users query questions on a set of retrieved papers, we explore how curated questions from domain-experts can be leveraged to prefill knowledge graph input templates for users. This complements the KG vision by lowering the barrier to entry and scaling up content acquisition. Embedding and indexing the extracted information in separate fields could lead to improved semantic search, by enabling users to search for papers with similar research questions or algorithms.
% The first sentence is not really clear and would be better rephrased. The first sentence also can fit as an intro sentence to Section 3.1

\subsection{LLMs for Information Extraction in Science}
In-context learning with LLMs describes the ability of these models to solve problems that they have not explicitly been trained on, by just giving the model an abstract description of the problem (\textbf{Zero-Shot}) or several examples (\textbf{Few-Shot}) in their input context. These approaches were first popularized with LLMs like GPT-3 \cite{brown2020language} and enable their use in domains where limited, or no training data is available. Recent LLMs such as the Google Gemini series\footnote{\url{https://web.archive.org/web/20250607225206/https://blog.google/technology/ai/google-gemini-next-generation-model-february-2024/}} or OpenAIs GPT-4.1\footnote{\url{https://web.archive.org/web/20250612080402/https://openai.com/index/gpt-4-1/}} have pushed the size of the available context up to 1 million input tokens. Which makes it feasible to extract information from large text sources in a single step.
These properties can be used in retrieval augmented generation (RAG) \cite{shuster2021retrieval} systems that ground the knowledge of these models in relevant texts. Systems such as CORE-GPT have shown the usefulness of such approaches in questions answering across multiple scientific domains \cite{pride2023core}.

In our work, we explore both zero- and few-shot learning for extracting predefined semantic information from scientific texts, that can then be used to facilitate scientific knowledge discovery and publication workflows. 

\section{Methodology: LLM-Based Concept Extraction}
\begin{figure}[htbp]
  \centering
  \includegraphics[width=.7\linewidth, alt={A diagram illustrating an in-context learning setup for an LLM. A user sends a 'Question & Paper' to a central Large Language Model (LLM). The LLM's processing is guided by two additional inputs: a set of 'Instructions' and a collection of 'Examples' (representing few-shot learning). After processing, the LLM returns an 'Answer' to the user.}]{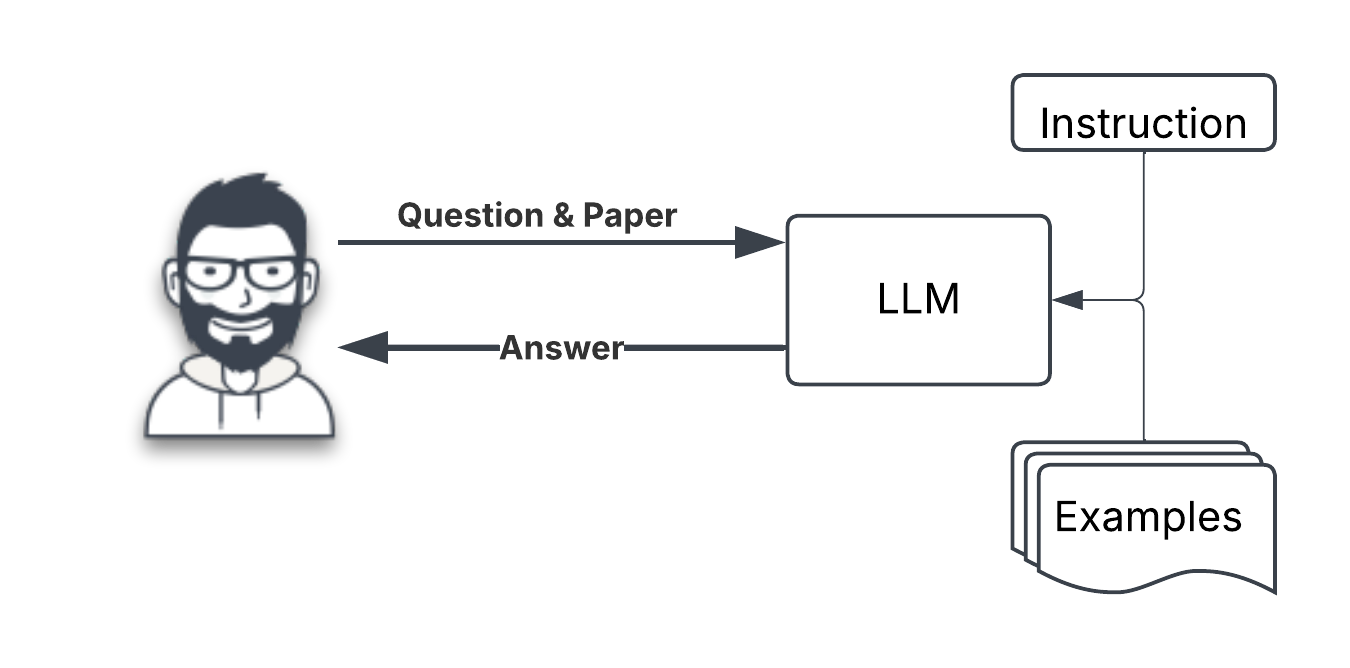}
  \caption{LLM-based demo extraction pipeline.}
  \label{fig:pipeline}
\end{figure}

Our system uses an LLM to extract semantic information from scientific documents. The demo UI allows a user to upload a paper and pose predefined or custom questions. The LLM then processes the document and a prompt to identify and return relevant information or synthesized answers (Figure \ref{fig:pipeline}).

\subsection{In-context Learning for Rapid Domain Adaptation}
We leverage the in-context learning of modern LLMs for rapid domain adaptation. By providing instructions and a few examples to guide extraction, our method avoids the need for the extensive, domain-specific datasets required by traditional supervised techniques.

In our evaluations we tested two modes: A \textbf{few-shot} mode where we supplied three in-domain examples each consisting of 1. the full text of the document; 2. the domain-specific information extraction questions, 3. the instructions, 4. the manually crafted ideal answer from the document. In addition, we tested a \textbf{zero-shot} mode where we only supplied the instruction and the full text of the PDF.

The \textbf{zero-shot} mode formed our baseline for evaluation, but could also be used to offer the flexibility to the user to pose their own extraction questions against a document or a set of retrieved documents. The \textbf{few-shot} mode aligns the model with the style of the manual annotators, overcoming limited or missing instructions in the way the question was posed. This mode could be used in settings where predefined questions and predictable answer formats are important, for example when prefilling forms for later knowledge graph mapping.

We tested four different models: Qwen 2.5 72B instruct \cite{qwen2025qwen25technicalreport}, Llama 3.3 70B instruct\footnote{\url{https://www.llama.com/docs/model-cards-and-prompt-formats/llama3\_3/}} \cite{grattafiori2024llama3herdmodels}, Gemini 1.5 Flash 002, Gemini 1.5 Flash 8B 001 \cite{geminiteam2024gemini15unlockingmultimodal}. Llama and Qwen were accessed via \url{openrouter.ai} while the Gemini models were accessed via the official Google API.

The instructions used chain of thought prompting \cite{10.5555/3600270.3602070} to generate a \textbf{reasoning} beside relevant \textbf{context} and the final \textbf{answer}.

The exact zero-shot prompt can be seen in Listing~\ref{lst:zero-shot-prompt}.
\lstdefinestyle{promptstyle}{
  basicstyle=\ttfamily\small\color{blue}, % Use monospaced blue text
  breaklines=true,                        % Enable line breaking
  postbreak=\mbox{\textcolor{blue}{$\hookrightarrow$}\space}, % Symbol at line break
  frame=single,                           % Optional: box around the prompt
  backgroundcolor=\color{gray!10},       % Optional: light gray background
  captionpos=b                            % Caption position
}
\begin{lstlisting}[caption={Zero-shot prompt example in Python}, label={lst:zero-shot-prompt}, 
  basicstyle=\ttfamily\small, breaklines=true, breakindent=0pt, frame=single]
Extract the information answering the following question from the text:
Question: ```{question}```
Text: ```{text}```
Return a JSON object in the following format: 
{{ "reasoning": "<think step by step and write down your reasoning>", 
"context": "<contains all relevant context from the text>", 
"answer": "<one concise answer to the question for example: yes/no/none, or a word or multiple words>" 
}}
Try to be concise and limit your reasoning, answer, and the extracted context to max 500 words.
\end{lstlisting}

\subsection{Dataset and Domain}
The initial development and a preliminary evaluation were carried out on a corpus of 122 scientific papers from the Business Process Management (BPM) conferences (2019–2023)\footnote{\url{https://bpm-conference.org/conferences/}}. This domain was selected due to the availability of domain experts who are actively constructing a knowledge graph in this area. Key concepts were manually annotated in the papers to establish a gold standard for evaluating extraction performance.

\subsection{Extraction Example}
An example of the extraction process for the \textbf{Target Concept} of a "Research Question" would involve posing the \textbf{Query}: "What is the explicitly stated research question for the paper?". The system might then return an \textbf{Example Extracted Answer} like: "How to decide which processes need to be analyzed in detail to determine if changes are necessary."

This methodology is highly flexible, allowing us to target a wide array of semantic information within scientific texts with high adaptability, as only limited domain expert involvement is needed to create a few examples for each information item.

\subsection{Demo System}
To showcase the ability of the tool and collect initial user feedback, we built a demo UI using the Gradio framework\cite{abid2019gradiohasslefreesharingtesting} around our approach. The demo system is available online\footnote{\url{https://demo-d3.nfdixcs.org/}}(user:demo, pw:demo) and the source code for this system is available on GitHub\footnote{\url{https://github.com/SamyAteia/nfdixcs-d3-knowledge-extraction-demo}}.

\subsection{User Feedback}
In a pilot user study, we collected initial user-feedback with the demo system through a questionnaire after instructing a panel of users to choose one paper from a selection of business processing domain papers, upload it to the tool and select any questions that they were interested in.

At a separate workshop with around 30 participants from different computer-science fields, we collected user-stories that they would like to be solved by the offered and demonstrated technology.

\section{Results}
We evaluated model performance against a gold-standard dataset of 122 manually annotated papers. We used three metrics based on the target type:
\begin{enumerate}
    \item For categorical targets (15 targets, 1121 annotations), we used \textbf{ExactAcc}, a token-set accuracy based on a Jaccard similarity with a threshold of 0.8 \cite{mann2016empirical,schmidt2025data}.
    \item For binary targets (13 targets, 984 annotations), we used the macro-averaged F1-score (\textbf{BinF1}).
    \item For free-text targets (4 targets, 488 annotations), we assessed semantic equivalence with the F1-score from \textbf{BERTScore} \cite{zhang2019bertscore}.
\end{enumerate}
The Overall score in Table~\ref{tab:model-comp} is the unweighted arithmetic mean of these three metrics.

\begin{table}[h!]
\centering
\caption{Model comparison on the paper-coding benchmark (higher = better)}
\label{tab:model-comp}
\begin{tabular}{|l|c|c|c|c|}
\toprule
\textbf{Model} & \textbf{ExactAcc} & \textbf{BinF1} & \textbf{BERT\_F1} & \textbf{Overall} \\
\midrule
qwen-2.5-72b (3-shot) & \textbf{0.249} & \textbf{0.594} & 0.863 & \textbf{0.569} \\
qwen-2.5-72b (0-shot) & 0.219 & 0.514 & 0.887 & 0.540 \\
llama-3.3-70b (0-shot) & 0.212 & 0.556 & 0.877 & 0.548 \\
gemini-1.5-flash-002 (3-shot) & 0.246 & 0.330 & 0.893 & 0.490 \\
gemini-1.5-flash-002 (0-shot) & 0.183 & 0.390 & 0.883 & 0.486 \\
gemini-1.5-flash-8b-001 (0-shot) & 0.171 & 0.345 & 0.881 & 0.466 \\
gemini-1.5-flash-8b-001 (3-shot) & 0.180 & 0.148 & \textbf{0.897} & 0.408 \\
\bottomrule
\end{tabular}
\end{table}

BERT\_F1 scores near 0.90 indicate strong semantic alignment on free-text fields, whereas binary indicators show moderate performance (best BinF1 = 0.59) and exact categorical extraction remains limited (ExactAcc < 0.25).

\subsection{User Study \& Workshop}
Feedback from our pilot user study on the prototype demo UI was positive (88\% satisfaction with extracted concepts), indicating the potential utility of the approach.
While some feedback was UI related (hiding advanced configuration like few-shot examples, wanting more expert configuration), a main point was that the traceability of the extracted information should be improved.

In a separate workshop with around 30 computer-science researchers, we collected 56 user-stories. 38 of these focused on the task of literature research and comparison, 8 on assistance while writing papers, 3 on support in the review process, 3 were directed towards software development and 4 were unique. Overall, it became clear that the users want to go beyond just extracting concepts from one specific paper and compare the extracted information from multiple papers instead. The full categorized list is available in our repository\footnote{\url{https://github.com/SamyAteia/nfdixcs-d3-knowledge-extraction-demo}}.

\section{Discussion}
The results of our technical evaluation and user-studies, while preliminary, provide valuable direction for the development of our future services.

Few-shot examples seemed to improve the performance of the models in tasks where specific categorial answers were needed and on the free-text extractions measured by BERTScore. But on the binary classification task, the performance decreased. This could be explained by a class bias introduced via the few-shot examples, while on the textual extractions the examples might have informed the model better about the expected format of the answers.

Using the full-text of documents in few-shot examples is costly and potentially increases noise. We are working on exploring the impact of more and  shorter examples and selecting ideal examples for specific extraction target types.

The open-weight models Qwen 2.5-72B-Instruct and Llama 3.3-70B-Instruct seemed to perform better than the commercial models that we tested. For the Gemini 1.5-flash-8b model, this is most likely explained by the difference in size. The size of the normal Gemini 1.5-flash model is unknown but given our results and the cost and speed we suspect that it is also smaller than the 70 billion parameter models that we compared them to.

From the feedback that we collected through our pilot user study and the discussion in a later workshop, it became clear that there is a need for better transparency and traceability. Ideally, highlighting the text passages that inform an extracted information items in the source document.

Even though there are commercial services available that are similar to our tool, our contribution can inform researchers and professionals that want to offer customizable domain-specific services to their users. We demonstrate the feasibility of in-context learning and open-weights models for these use-cases and publish our code to boost independent development of transparent services. 

\section{Conclusion and Ongoing Work}
We confirmed the potential of current LLMs to summarize and extract domain-specific information from scientific text. Through in-context learning, these models can be quickly adapted to specific scientific domains and facilitate the transfer of expert knowledge between researchers by highlighting and comparing key aspects of their work. 

Through our user-centric approach, we collected valuable feedback and user-stories that can guide the development of current and future services. Notable transparency and the need that services enable the user to verify LLM generated output by tracing summarized information back to the source text.  

Our ongoing work will focus on exploring embedding-based retrieval on the extracted structured information, therefore overcoming the arbitrary chunking issue that limits semantic relevance in vector search. We're also exploring how our approach can be integrated in the publishing process, prefilling templates for knowledge graph mapping e.g., for ORKG. Making it easier for authors to fill out forms that facilitate the discoverability of their work.

Overall, our work highlights the potential of LLMs to improve the publishing process and discoverability of scientific information in digital libraries and beyond. Its main contribution is demonstrating the practical integration of these models into a user-focused, open-source system designed to tackle real-world challenges, rather than proposing a novel extraction algorithm itself.

\begin{credits}
\subsubsection{\ackname}
We thank the anonymous reviewers for their valuable feedback. This work is funded by the German Research Foundation (DFG) as part of the NFDIxCS consortium (Grant number: 501930651).

\subsubsection{\discintname}
The authors have no competing interests to declare that are relevant to the content of this article.
\end{credits}
\bibliographystyle{splncs04}
\bibliography{bibliography}

\end{document}